\documentclass[a4paper, amsfonts, amssymb, amsmath, reprint, showkeys, nofootinbib, twoside]{revtex4-1}

\usepackage{dcolumn}% Align table columns on decimal point
\usepackage{bm}% bold math

\usepackage{graphicx}
\usepackage{amsmath}
\usepackage{amsfonts}
\usepackage{braket}
\usepackage{amssymb}
\usepackage{enumerate}
\usepackage{multirow}
\usepackage{color}
\usepackage{url}
\usepackage{bbm}
\usepackage{caption}
\usepackage{subcaption}
\usepackage{mathtools}
\usepackage{bbold}
\usepackage{hyperref}
\hypersetup{
  colorlinks   = true, 
  urlcolor     = blue,
  linkcolor    = blue, 
  citecolor   = red
}
\newcommand{\norm}[1]{\left\lVert#1\right\rVert}

\newtheorem{definition}{Definition}[section]

\DeclareMathOperator{\Tr}{Tr}

\begin{document}

\preprint{APS/123-QED}

\title{Robust Quantum Gate Complexity: Foundations}

\author{Johannes Aspman}
\author{Vyacheslav Kungurtsev}
\author{Jakub Marecek}
\affiliation{Department of Computer Science, Czech Technical University in Prague,\\ Karlovo nam. 13, Prague 2, Czech Republic}
\author{All authors contributed equally.}

\maketitle

\section{Introduction}
Optimal control of closed quantum systems is a well studied geometrically elegant set of computational theory and techniques that have proven pivotal in the implementation and understanding of quantum computers. The design of a circuit itself corresponds to an optimal control problem of choosing the appropriate set of gates (which appear as control operands) in order to steer a qubit from an initial, easily prepared state, to one that is informative to the user in some sense, for e.g., an oracle whose evaluation is part of the circuit. However, contemporary devices are known to be noisy, and it is not certain that a circuit will behave as intended. Yet, although the computational tools exist in broader optimal control theory, robustness of adequate operation of a quantum control system with respect to uncertainty and errors has not yet been broadly studied in the literature. Some previous works have addressed the question of controllability of open quantum system, see for example \cite{altafini2004coherent, wu2007controllability,jover2017open}, while the recent work \cite{kallush2022controlling} does study the optimal control problem with some interesting results. In this paper, we propose a different approach than \cite{kallush2022controlling}, making use of geometric interpretations of the optimal control problem. To this end, we present the appropriate problem definitions of robustness in the context of quantum control, focusing on its broader implications for gate complexity. 

In a series of seminal papers~\cite{nielsen2006geometric,nielsen2006optimal,nielsen2006quantum,dowling2008geometry}, Nielsen et al. established a formal correspondence between the time optimal closed quantum control problem, the gate complexity of a quantum circuit, and the geodesic path on a manifold corresponding to the geometry of quantum operators, with the metric appropriately informed by the properties of the available gates.

This line of work has developed towards increased sophistication and broader applicability. In~\cite{dowling2008geometry}, the authors formally develop the control/geometry correspondence associated with the metric's Christoffel symbols and parallel transport.
Simple analytically derived examples were studied in \cite{PhysRevA.78.032327,PhysRevD.100.046020}. See also \cite{khaneja2002sub} for an application to NMR and \cite{PhysRevD.95.045010,Jefferson2017} for connections to high-energy physics. 

Most recently, \cite{brown2018second,Haferkamp2022} have used this reasoning to show the linear growth of circuit complexity, in terms of the number of Haar-random two-qubit quantum gates required to implement a unitary.
While the original metric of Nielsen \cite{PhysRevA.78.032327} is known to be non-smooth \cite{Bulchandani2021} (sometimes known as the ``cliff metric''), there also exist smooth variants \cite{brown2021quantum}. 
For some discussion of recent developments, see~\cite{Bulchandani2021,auzzi2021geometry,PhysRevD.100.046020,brown2021quantum}. 
Near term devices are all expected, however, to exhibit significant noise with any circuit of reasonable depth. Thus there is always some probability that the desired state will not be achieved, and only proximity with high probability could be a reasonable target. The first work to acknowledge this point formally~\cite{tibbetts2012exploring}, studied tradeoffs associated with inexact satisfaction of the time optimal control. They computed a pareto front of cost and gate fidelity demonstrating the relative effort required to ensure minimal gate error. 

Control of open quantum systems have been discussed scarcely in the literature, see for example \cite{altafini2004coherent, wu2007controllability,jover2017open, kallush2022controlling}. Most of this literature deals with the question of controllability. In \cite{kallush2022controlling}, however, the question of optimal control for open quantum systems is considered. There, the authors consider an iterative approach for the optimal control of Lindbladian open quantum systems. Our proposal is rather different as we generally do not require an iterative solution of the dynamics at each time step.

The standard time optimal control associated with the gate complexity problem~\cite{nielsen2006geometric,nielsen2006optimal,nielsen2006quantum,dowling2008geometry} minimizes the time required to synthesize a unitary target $\bar{U}$ under Schr{\"o}dinger dynamics,  i.e.,
\begin{align}
\min\limits_{\{h_j(t)\}\in \mathcal{H}} \;\;& T \label{eq:ocpdynamics:a}\\
    \text{subject to }\frac{dU}{dt}&=-iH(t)U(t) \label{eq:ocpdynamics:b}\\
     U(T) &\coloneqq \bar{U} \label{eq:ocpdynamics:c}\\
     H(t)&\coloneqq \sum\limits_{j=1}^m h_j(t) H_j \label{eq:ocpdynamics:d}
\end{align}
where $h_j(t)$ are control variables. They correspond to implementing a mixture of Hamiltonian operators $\{H_j\}$ available to the controller.

We depart from this standard formulation and introduce error in the operator application, in the form of an unknown time dependent switching term selecting between the intended operation and some other operator. Formally, 
\begin{equation}\label{eq:pertH}
H_j(t) = \alpha_j(t)\hat{H}_j+(1-\alpha_j(t))\breve{H}_j
\end{equation}
where $\hat{H}_j$ is the desired Hamiltonian and $\breve{H}_j$ the erroneous one. In this first work we consider that $\breve{H}_j$ is Hermitian, thus preserving the geometry of the control as being on the manifold of unitary matrices. 

With some probability, any intended trajectory could be thwarted by the presence of errors, i.e., there shall be some $t_a,t_b$ with $\alpha_j(t)=0$ for $t_a<t<t_b$. This necessitates considering alternative, approximate notions of the time optimal control and the associated notions of gate complexity and geometric distance minimizing paths. 

First, a statistical agglomeration must be chosen. Classically, the expectation is applied on the cost function in the optimal control problem, and now the control to minimize the expected cost is sought. Then, since analytic solutions are rarely tractable for these problems, some sampling approach must be used with the numerical optimization to form a well defined finite-dimensional problem. In numerical optimal control, the standard technique is to introduce \emph{scenarios} whereby a \emph{sample average approximation} of the objective is constructed and defined to be the target.
\begin{figure*}
\includegraphics[scale=0.45]{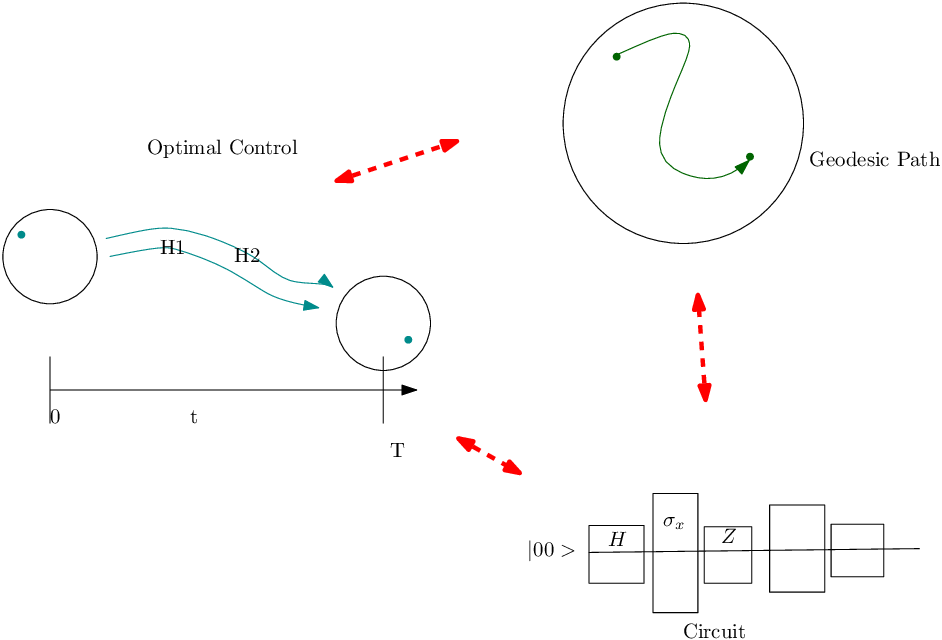} \quad  
\includegraphics[scale=0.45]{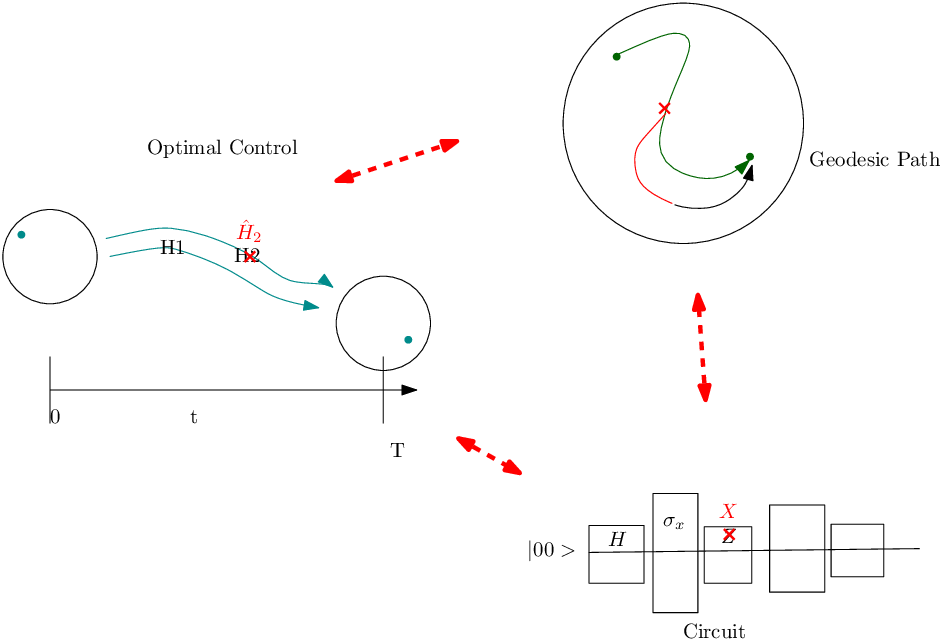}
\caption{Left: Schematic of the correspondence described in the papers of Nielsen et al. \cite{nielsen2006geometric,nielsen2006optimal,nielsen2006quantum,dowling2008geometry}
Right: 
Schematic of the correspondence when subject to gate errors. In this case a switch of the application of $H_2$ corresponds to changing $Z$ to $X$ gate and an unintended veering of the trajectory on the manifold that must be subsequently corrected for.}
%\caption{}
%\end{figure}
\end{figure*}
In this brief paper, we formulate the robust time optimal control problem, indicate appropriate methods and procedure to solve it, and derive the corresponding perturbative notions for the gate complexity and geodesics. Thus, we introduce considerations of gate errors into the theoretical framework of geometric quantum control in terms of how the control relates to gate complexity and path geometry.

\begin{figure}
\includegraphics[scale=0.35]{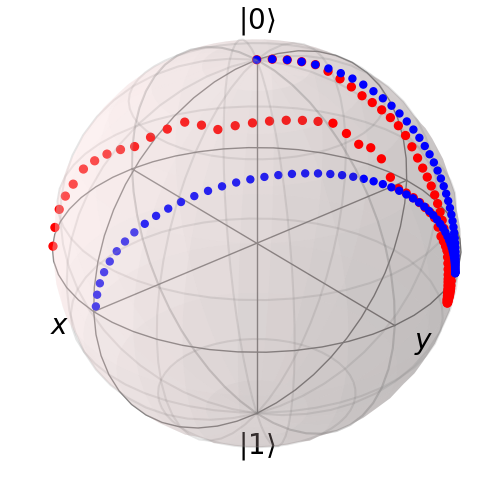}
\caption{Schematic of the effect of noise on quantum control. Here we are using the Hamiltonian defined by $\hat{H}_x=\sigma_x$, $\hat{H}_y=\sigma_y$, and $\breve{H}_x=\breve{H}_y=\sigma_x$. The target unitary is the Hadamard gate. The blue dots indicate the state transformation under the controlled process without noise, starting from the initial state $|0\rangle$, while the red dots indicate the evolution when the dynamics are controlled without accounting for noise.}
\end{figure}

\section{The Robust Optimal Control}

The aforementioned literature established formal isomorphisms between the time optimal control problem, the gate complexity of a circuit, and the minimal distance computed between an initial and target state. See the supplementary material for the formal technical details. In the presence of gate errors, this picture needs to be augmented with an appropriate problem formulation that targets a statistic, with respect to the noise, of the optimal time, and correspondingly, the gate complexity and geodesic distance. We shall describe how the standard application of robust scenario based stochastic optimal control on this construction is formed and the kinds of problems and possible algorithms that could be used to solve them. 

Fortunately, when it comes to robust optimal control, there are formulations that exist in the literature, with the most well established ones being based around solving sample average approximations of the problem. In the stochastic programming literature, the operative terminology is using a set of \emph{scenarios}, and formulating the discretized problem wherein it is enforced that a control action cannot be dependent on a future realization.

\subsection{Modeling gate errors and noise}
Gate error noise corresponds to an application of an operator that is distinct from the intended one. We expect that the tendency for a gate to yield error is auto-correlated across time, so if correctly applied at time $t$ it is likely to be correctly applied at time $t+dt$. Otherwise $\alpha_j(t)$ in~\eqref{eq:pertH} as a function of $t$ could have an infinite number of discontinuous points within any finite time interval, potentially causing regularity issues in studying the OCP.

The appropriate noise process then takes $\alpha_j(t)$ as \emph{bang-bang}: $\alpha_j(t) = 0$ for some $t\in[0,T]$ and $1$ for the complement on $[0, T]$. When $\alpha_j(t)$ jumps in that interval is random. This corresponds to a situation wherein we can apply a gate multiple times, with an error appearing at some instances and others not, without the time of the error appearing being known a priori. 
We can consider the possible realizations of $\alpha(t)$, denoted by $\mathcal{A}$ to be the set of $\{0,1\}$ valued maps with $\text{meas}(\{t:\alpha_j(t)=0\})\ge T/a$ for some $0<a<T$. To avoid certain degeneracies that would come up otherwise, we consider that $a\ge T/2$.

A natural parametrization is as a Poisson point process, i.e., with exponentially distributed jump times, with $\alpha_j(0)=1$ and,
\[
\mathbb{P}(\alpha_j(t)=0,\,\alpha_j(s)=1,\,s<t) = 1-e^{-\lambda_e t}
\]
then,
\begin{equation*}
\begin{aligned}
    &\mathbb{P}(\alpha_j(t)=1,\,\alpha_j(s)=0,\,t_0\le s<t,\, \alpha_j(r)=1,\,r<s) \\
    &= 1-e^{-\lambda_c t}
\end{aligned}
\end{equation*}
where we can consider asymmetry and the time to balance the lack of errors as more likely/common, i.e., $\lambda_c<\lambda_e$. 

\subsection{The Robust Control Formulation}
With the change in the model to incorporate errors, we must now reformulate the optimal control problem. As noted, since the outcome depends on the stochastic realization, the original problem is now ill-defined, and for every equation appearing in~\eqref{eq:ocpdynamics:a}-\eqref{eq:ocpdynamics:d} which now would appear as probability-valued, some operation aggregating the distribution must be performed. 
\begin{align}
\min\limits_{\{h_j(t)\}\in\mathcal{H}} \;\;& \mathcal{R}_{\xi}(T) \label{eq:ocpdynamicsstoch:a}\\
    \text{subject to }&\frac{dU(\cdot;\xi)}{dt}=-iH(t;\xi)U(t;\xi) \label{eq:ocpdynamicsstoch:b}\\
    & \mathbb{P}_{\xi}\left[\|U(T) - \bar{U}\|_{\infty} \le \eta\right]\ge 1-\beta \label{eq:ocpdynamicsstoch:c}\\
    &H(t;\xi):=\sum\limits_{j=1}^m h_j(t) H_j(t) \label{eq:ocpdynamicsstoch:d} \\
    & H_j(t;\xi) = \alpha_j(t;\xi)\hat{H}_j+(1-\alpha_j(t;\xi))\breve{H}_j, \label{eq:ocpdynamicsstoch:e}\\
    & \alpha_j(t;\xi) = \mathcal{L}(\mathcal{P})
    \label{eq:ocpdynamicsstoch:f}
\end{align}
Throughout we use $\xi$ to denote the randomness in the process. A possible given value of $\xi$ corresponds to a particular possible trajectory of $\alpha_j(t)$ for all $j$. $\mathcal{L}$ is the law of the stochastic process $\mathcal{P}$ described above governing $\alpha_j$, that is the meaning of ~\eqref{eq:ocpdynamicsstoch:f}. 

While the controls themselves $h_j(t)$ are deterministic, their effect on the state is mediated by the $\{\alpha_j(t;\xi)\}$ as expressed in the expression for the actual operator applied given in~\eqref{eq:ocpdynamicsstoch:e}. In~\eqref{eq:ocpdynamicsstoch:b}, we index the state by the stochastic realization $\xi$. Thus we have a random ODE, with an ODE defined for every possible value of $\xi$.

In the objective~\eqref{eq:ocpdynamicsstoch:a}, the time has been replaced by some operator $\mathcal{R}_{\xi}$ of the time. In robust OCP this operator can be a statistical moment or a coherent risk measure (e.g., the expectation $\mathbb{E}[T]$, or the conditional value at risk, CVaR).

Finally we remark on~\eqref{eq:ocpdynamicsstoch:c}. This is a \emph{chance constraint}, respecting that we are not certain to arrive at the target $\bar{U}$, and so we can at best aim to reach a neighborhood, defined by some $\eta>0$, with some high probability.
It can be observed that this expression resembles the probably approximately correct model of learning. In this sense, the robust optimization problem learns a gate set that aims to learn the procedure to form the desired unitary. The principled methodological introduction of statistical learning techniques for studying this problem would be a potentially interesting and useful topic for future work.

\subsection{Implications for Geometry and Complexity}
It turns out that one can write the expression for the action principle for the geodesic equations as a function of the controls, and thus one can obtain the equivalent equation for minimizing the time as by minimizing the corresponding distance on the manifold. Now, however, the distance is considered in a robust sense; we are taking an expectation or risk measure computation of it. 

The link to gate complexity can be seen to have functional correspondence to the robust optimal control problem both directly
and mediated through the geometry. For the first, with the understanding that the optimal control shall be discretized in time, one can immediately consider the total number of discretized intervals to be the total quantity of operators applied, and thus the gate complexity. As the total number of intervals is itself variable, indeed it is the quantity to be minimized in the cost function, this quantity has an exact correspondence to the gate complexity. The chance constraint carries over, and the gate complexity is one of high confidence, probability $1-\beta$ for some small $\beta$. 

The gate complexity becomes more transparently linked trough this paradigm by the geometry. In particular, the choice of a finite gate set can be used to inform the metric in regards to path distances as depending on the direction of manifold traversal. In the Appendix we present some metrics associated with certain gate sets in the literature. Thus an operator can represent a functional operation on a qubit in the OCP, a pushforward along a geodesic path with a certain distance traversed for the geodesic functional minimization problem, and a gate in a circuit to perform some intended computational operation with varying associated progress towards the ultimate circuit computation desired.

\subsection{Methods for solving the problem}

The Magnus expansion \cite{blanes2009magnus} is a powerful tool for approximation of solutions of ordinary differential equations appearing throughout mathematics and physics. This has been applied with a great level of success to various fields of physics, including closed quantum optimal control \cite{marecek2020quantum,bondar2022quantum}. On the other hand, when considering systems with inherent noise, stochastic differential equations are the appropriate model to use. The application of Magnus expansion techniques to stochastic differential equations has been limited, even though some literature exists \cite{wang2020magnus, wang2020new, yang2021class}. The theory of stochastic Magnus expansion should serve as an interesting approach for the problem of robust quantum control.

On the other hand, a natural approach to solving the problems described above is Sample Average Approximation (SAA). In SAA one performs a number of samples of the random variables in the problem, and computes a sample average of the resulting deterministic optimization problems. 
Clearly, one needs to sample the random variables associated with the noise process, and thus  trajectories. Associated with each realization is a path $U(t,\xi_l)$ and time $T_{\xi_l}$. The cost function could then be, in the case of $\mathcal{R}_{\xi} = \mathbb{E}_{\xi}$, recognizing that the state is $\xi$ dependent but not the control $h_j(t)$.
In addition, in order to present a problem amenable to practical computation, the space of control functions also needs to be discretized. Specifically, we discretize the time by $\Delta t$, i.e., set $t_0=0,\,t_1=\Delta t,\,t_2=2\Delta t$, etc. Given that with time optimal control, we do not know a priori how many time steps will be necessary, we incorporate an initial large value as a guess, noting that after any final time $T^{l}$, controls are not expected to influence the outcome for that particular trial.

\section{Discussion and Conclusion}
In this brief paper, we have discussed the time-optimal control problem for open quantum systems. We presented a robust control formulation for a model of a noisy open quantum system, where the noise is unitary, and discussed which methods and algorithms are equipped to solve this, in particular, we discussed applications of the Magnus expansion for stochastic differential equations as well as the Sample Average Approximation as means to numerically solve the control problems. 

In the Appendix we discuss some more details on the correspondence between geometry and gate complexity, and provide a theoretical bound on the gate complexity based on the corresponding derivation for the closed system by \cite{Brown:2022phc}.

\clearpage

\appendix

\section{Gate complexity and geometry}
Following Nielsen et al., we are interested in studying distances on the group manifold $SU(2^n)$ with respect to various example metrics. A set of bases for the tangent space of $SU(2^n)$ is provided by the generalized Pauli matrices $\sigma_I$, $I=1,\dots,2^{2n}-1$, given by traceless tensor products of the ordinary Pauli matrices and the identity matrix. As an example, the tangent space of the two qubit case, $n=2$, is spanned by the fifteen matrices
\begin{equation}
    \{\sigma_j\otimes \mathbbm{1},\, \mathbbm{1}\otimes \sigma_j,\,\sigma_j\otimes\sigma_k\},\qquad j,k=x,y,z.
\end{equation}
We further define the weight of an element in the generalized Pauli basis as the number of factors in the tensor product not equal to the identity. In the two qubit example this means that $\{\sigma_j\otimes \mathbbm{1},\, \mathbbm{1}\otimes \sigma_j\}$ has weight one, while $\{\sigma_j\otimes\sigma_k\}$ has weight two. 

\subsection{Gate complexity} 
Gate complexity (sometimes called circuit complexity) is a measure on how complicated it is to generate, or at least approximate, a certain state or unitary operator. As such, it is given by the minimal number of primitive gates needed in the circuit to approximate the requested state or unitary to within some error.

%Definition 3.2.5 in https://arxiv.org/pdf/1607.05256.pdf

\begin{definition}[Gate Complexity]
We let $\mathcal C^A_\epsilon(U)$ be the minimum size of a quantum circuit with gates $A$ that $\epsilon$-approximates $U$.
\end{definition}

To fully characterize the gate complexity of a given system we then need to pick a set of primitive gates together with a tolerance level $\epsilon$. From the Solovay-Kitaev theorem, we know that given any physically universal gate set, a target unitary, and accuracy $\epsilon$, the number of gates that are needed to approximate the target unitary to  the entrywise $\epsilon$ accuracy is $\log^{O(1)}(1/\epsilon)$. 

%See Corollary 1 in \cite{jia2022hay} for explicit examples of exponential circuit complexity.

One drawback of gate complexity, that led to the introduction of complexity geometry by Nielsen and collaborators, is that it is discontinuous. Since the circuit is discrete, there can be points in the space of unitaries that are arbitrarily close in the ordinary inner product sense, but exponentially far away in gate complexity. By considering a continuous time-optimal control problem, Nielsen and collaborators, \cite{nielsen2006geometric,nielsen2006optimal}, presented a framework of using optimal control theory, with its comprehensive literature of theoretical guarantees and algorithms, to make definitive statements regarding otherwise cumbersome analysis of gate complexity outright.

\subsection{Metrics}
Points on the manifold $SU(2^n)$ correspond to unitary matrices $U$ and we thus construct metrics by studying the distance to a nearby point $U+dU$. The most straightforward example is the ordinary inner-product, or Killing, metric
\begin{equation}\label{eq:innProdMetric}
    ds^2=\Tr[dU^\dagger dU]=\sum_{I,J}\Tr[idU U^\dagger \sigma_I]\delta_{IJ}\Tr[idU U^\dagger\sigma_J],
\end{equation}
where the trace has been normalized such that $\Tr[\mathbbm{1}]=1$. The approach of Nielsen et al. was to generalize this metric by introducing a penalty matrix $\mathcal{I}_{IJ}$ that encodes the idea that certain directions on $SU(2^n)$ might be more costly to move in than others. Here, $\mathcal{I}_{JK}$ is positive and symmetric. The resulting family of metrics is given by
\begin{equation}\label{eq:costmetric}
    ds^2=\sum_{I,J}\Tr[idU U^\dagger \sigma_I]\mathcal{I}_{IJ}\Tr[idU U^\dagger\sigma_J].
\end{equation}
An important aspect of these metrics is that they are right-invariant, meaning that it is invariant under $U\to UU_R$, while, in contrast, generally not being left-invariant. A metric that is both right- and left-invariant is called bi-invariant, and the ordinary inner-product metric of \eqref{eq:innProdMetric} is an example.

We will assume that $\mathcal{I}_{IJ}$ is diagonal with respect to the generalized Pauli basis, such that the metric becomes
\begin{equation}\label{eq:genMetric}
    ds^2=\sum_{I}\mathcal{I}_{II}\Tr[idU U^\dagger \sigma_I]^2.
\end{equation}
A natural choice for the metric is to penalize directions with high weight, i.e. making it easier to act on a small number of qubits. The penalty matrix is then only dependent on the weight of the basis element and we specify the metric completely by specifying  $\mathcal{I}_{II}\eqqcolon\mathcal{I}_k$ for $\sigma_I$ having weight $k$. Three such examples of metrics are studied in \cite{Brown:2022phc}:
\begin{itemize}\setlength\itemsep{1em}
    \item Firstly, we can consider the case of penalizing basis directions with weight higher than two. In other words we set $\mathcal{I}_{II}=\mathcal{I}_k=1$ whenever $\sigma_I$ has weight $k=1,2$, and $\mathcal{I}_{II}=\mathcal{I}_k=k$ when $\sigma_I$ has weight $k>2$. This is often referred to as the \emph{cliff metric}, and corresponds to the case already studied in the original works by Nielsen et al. \cite{nielsen2006geometric,nielsen2006optimal,nielsen2006quantum,dowling2008geometry, PhysRevA.78.032327}.
    
    \item Secondly, we can consider adding a penalty that depends on how often a certain weight in the basis appears. For the generalized Pauli basis, there are exactly $\begin{pmatrix}n\\k\end{pmatrix}3^k$ elements having weight $k$. We can thus consider the \emph{binomial metric} as having $\mathcal{I}_k(\alpha)=\left(\begin{pmatrix}n\\k\end{pmatrix}3^k\right)^\alpha$, for some parameter $\alpha$ \cite{Brown:2022phc}. 
    
    \item Finally, we can consider simply increasing the penalty with increasing weight. This gives the \emph{exponential metric}, $\mathcal{I}_k(x)=x^{2k}$ of \cite{Brown:2022phc}. 
    
\end{itemize}
Of course, in the single qubit case we only have weight one operators and the above is not applicable. A natural approach is then to pick out one particular direction, say $\mathcal{I}_{zz}$, to penalize, setting $\mathcal{I}_{xx}=\mathcal{I}_{yy}=1$ \cite{PhysRevA.78.032327,PhysRevD.100.046020}. Similarly, for the two qubit case we simply penalize the directions with weight two (corresponding to the operators $\{\sigma_j\otimes\sigma_k\}$), setting $\mathcal{I}_{II}=1$ for the other directions.

\subsection{Geodesics and optimal control}
The important insight of Nielsen et al. was to equate the optimal value of the cost function with a right invariant metric on $SU(2^n)$,
\begin{equation}
    c(H,T)=\int_0^T\sqrt{g(H(t), H(t))}dt,
\end{equation}
where $g(\cdot,\cdot)$ denotes a specific metric. This is equal to the length of the curve mapped out by the Schr{\"o}dinger evolution. When we do not have any noise, we can express the Hamiltonian in the generalized Pauli basis as
\begin{equation}
    H(t)=\sum_I h_I(t)\sigma_I.
\end{equation}
For the general family of metrics, \eqref{eq:genMetric}, we then find 
\begin{equation}
    g(H,H)=\sum_I \mathcal{I}_{II}h_I^2.
\end{equation}
We now define the operator complexity by
\begin{equation}
    C(U)=\inf_{H,T}c(H,T),
\end{equation}
where we take the infimum over the time $T$ and over the control Hamiltonians that synthesize $\bar U$. Looking for minimal curves on manifolds is of course the same as looking for minimal length geodesics with respect to the given metric. Thus, in the case wherein the control spans the space, which corresponds to the generality of parametrizing an arbitrary curve, the cost can be identically computed by solving the optimal control problem as solving the geodesic equations. 

\subsection{Geodesics with Unitary Noise}

If we assume that the noise is Hermitian, such that the geometry is preserved. We can then expand $\breve{H}$ in the generalized Pauli basis,
\begin{equation}
    \breve H_J=\sum_I M_{IJ}\sigma_I,
\end{equation}
for some constants, or possibly time dependent, $M_{IJ}$. We now find that the ``difficulty" of applying $H$ is given by
\begin{equation}\label{eq:costPert}
    \begin{aligned}
    g(H,H)=&
    \sum_I \mathcal{I}_{II}(\alpha_I h_I)^2\\
    &+2\sum_{I}\sum_{J\neq I}\mathcal{I}_{II}\alpha_I(1-\alpha_J)h_Ih_JM_{IJ}\\
    &+\sum_{I,J,K}\mathcal{I}_{II}(1-\alpha_I)(1-\alpha_J)h_Ih_J M_{KI}M_{KJ},
    \end{aligned}
\end{equation}
and the length of the trajectory is given by
\begin{equation}
    c(H,T)=\int_0^T\sqrt{g(H(t),H(t))}dt.
\end{equation}

\section{Bounding gate complexity}
To bound the gate complexity, we follow Brown \cite{Brown:2022phc} and make three approximations of our path. Our target unitary $\bar U$ is given by
\begin{equation}
    \bar U=\mathcal{P}\exp\left[ i\int dt H(t)\right].
\end{equation}
We now make three approximations
\begin{enumerate}\setlength\itemsep{1em}
    \item We first cut out the most expensive components by introducing a cut-off $\bar{\mathcal{I}}$,
    \begin{equation}
        H(t)\to  H_P(t)=\sum_{\substack{I\\ \mathcal{I}_{II}<\bar{\mathcal{I}}}}h_I(t)H_I(t).
    \end{equation}
    
    \item Next, we divide the path into $S$ equal steps of inner-product length $\delta$ and take the average Hamiltonian on each step,
    \begin{equation}
        \mathcal{P}\exp\left[ i\int_{T}^{T+\delta} dt H(t)\right]\to \exp\left[ i\int_{T}^{T+\delta} dt H(t)\right],
    \end{equation}
    for each time possible time $T$. This is similar to the popular GRAPE algorithm \cite{khaneja2005optimal}.
    
    \item Finally, we ``trotterize", meaning that we apply each term in the expansion of the Hamiltonian sequentially,
    \begin{equation}
        \exp\left[i\int dt\sum_I h_I(t)H_I(t)\right]\to\prod_{I} \exp\left[i\int dt h_I(t)H_I(t)\right].
    \end{equation}
\end{enumerate}
Each of these steps introduces some error that we need to bound. To this end, we introduce the notation $s(\cdot,\cdot)$ for the Killing metric. Note that the relation to $g(\cdot,\cdot)$ of say Eq. \eqref{eq:costPert} is just to set $\mathcal{I}_{IJ}=\delta_{IJ}$. 

\begin{enumerate}
    \item In the Killing metric, the error from pruning is bounded by realising
    \begin{equation}
        \begin{aligned}
            s&\left(\mathcal{P}e^{i\int H(t)dt},\mathcal{P}e^{i\int H_P(t)dt}\right)\\
            \leq& \int dt\sqrt{\text{Tr}[(H(t)-H_P(t))^2]}\\
            =&\int dt\sqrt{\text{Tr}[(\sum_{I:\mathcal{I}_{II}\geq \bar{\mathcal{I}}}h_IH_I(t))^2]}.
        \end{aligned}
    \end{equation}
    We recognize the factor under the square root as \eqref{eq:costPert} with $\mathcal{I}_{II}=1$ and where the summand is restricted due to the pruning. Denoting $\Gamma^2\coloneqq g(H,H)$ we have
    \begin{equation}
        \Gamma^2=\sum_I \mathcal{I}_{II}(\dots)\geq \bar{\mathcal{I}}\sum_{I: \mathcal{I}_{II}\geq\bar{\mathcal{I}}}(\dots),
    \end{equation}
    where the dots represent everything within the sum of \eqref{eq:costPert}. This gives us
    \begin{equation}
        s\left(\mathcal{P}e^{i\int H(t)dt},\mathcal{P}e^{i\int H_p(t)dt}\right)\leq \int dt \frac{\Gamma}{\sqrt{\bar{\mathcal{I}}}}.
    \end{equation}

    \item To find the Killing error from averaging, we first note that it is a known fact that for any operators $A$ and $B$ we have \cite{Brown:2022phc}
    \begin{equation}
        \norm{AB}_F\leq \norm{A}_F\norm{B}_{op.},
    \end{equation}
    where we normalise the Frobenius norm such that $\norm{H(t)}\leq 1$. Brown also shows that, for a Hamiltonian $H$ with $N$ non-zero generalized Pauli matrices, we have the inequality
    \begin{equation}
        \norm{H}_{op.}\leq \sqrt{N}\norm{H}_F.
    \end{equation}
    Combining these two results we have
    \begin{equation}
        \norm{H_1\cdots H_m}_F\leq N^{\frac{m-1}{2}}\norm{H_1}_F\cdots \norm{H_m}_F.
    \end{equation}
    Now, to bound the error from averaging we use the Dyson expansion. For one time interval $t\in [0,\delta]$ we write $H_{av.}\coloneqq \delta^{-1}\int_0^\delta dtH_P(t)$, and we thus have
 \begin{equation}
        \resizebox{0.5\textwidth}{!}{$\begin{aligned}
            &\norm{\exp[i\int_0^\delta dtH_P(t)]-\mathcal{P}\exp[i\int_0^\delta dtH_P(t)]}_F\\
            =&\norm{\sum_{m=2}^\infty \frac{(-iH_{av.} \delta)^m}{m!}-\sum_{m=2}^\infty(-i)^m\int_0^\delta dt_1\int_0^{t_1}dt_2\cdots \int_0^{t_{m-1}}dt_mH_P(t_1)\cdots H_P(t_m)}_F\\
            \leq&\sum_{m=2}^\infty \frac{\norm{(-iH_{av.}\delta)^m}_F}{m!}+\sum_{m=2}^\infty \int_0^\delta dt_1\cdots \int_0^{t_{m-1}}dt_m\norm{H_P(t_1)\cdots H_P(t_m)}_F\\
            \leq&2\sum_{m=2}^\infty\frac{N^{\frac{m-1}{2}}\delta^m}{m!}\\
            \leq& \frac{2\left(e^{\sqrt{N}\delta}-1-\sqrt{N}\delta\right)}{\sqrt{N}},
        \end{aligned}$}
    \end{equation}
where we used our normalisation $\norm{H}_F\leq 1$. If we now pick $\delta<N^{-\tfrac12}$ we find
\begin{equation}
    \norm{\exp[i\int_0^\delta dtH_P(t)]-\mathcal{P}\exp[i\int_0^\delta dtH_P(t)]}_F< 2\sqrt{N}\delta^2,
\end{equation}
which implies that the Killing error is bounded by
\begin{equation}
    s\left(\mathcal{P}\exp[i\int_0^\delta dtH_P(t)],\exp[i\int_0^\delta dt H_P(t)]\right)<\pi \sqrt{N}\delta^2,
\end{equation}
where we used that the Killing metric is bounded by the Frobenius norm \cite{Brown:2022phc}.

    \item Finally, the error from the trotterization has been actively studied in recent years, see for example \cite{layden2022first}. We consider the error
    \begin{equation}
        \begin{aligned}
        \text{err.}_{\text{trot.}}\coloneqq& \norm{e^{i\int_T^{T+\delta}dt\sum_I H_I}-\prod_{I=1}^N\left( e^{i\int_T^{T+\delta}dtH_I}\right)}\\
        =&\norm{\frac{1}{2}\sum_I\sum_{J<I}[H_I,H_J]\delta^2+\mathcal{O}(\delta^3)}.
        \end{aligned}
    \end{equation}
    In our case, when we have noisy control Hamiltonians, the consideration is a bit more subtle than in the closed quantum case of \cite{Brown:2022phc}, since we can have constructive interference 
\end{enumerate}

We will expand on point 3. above and finish bounding the gate complexity of the open quantum system similar to the discussion in  Section 3.2 of \cite{Brown:2022phc} in future work.

%\section{One-qubit example?}

\end{document}